\pdfoutput=1
\documentclass[runningheads]{llncs}

\usepackage[misc]{ifsym}
\usepackage{gensymb}
\usepackage{textcomp}
\usepackage{hyperref}
\usepackage[separate-uncertainty = true,round-mode=places,round-precision=2,
bracket-numbers = false,multi-part-units = single]{siunitx}
\usepackage{booktabs}
\usepackage[T1]{fontenc}
\usepackage{lmodern}

\usepackage{graphicx}
\graphicspath{{figures/}}

\begin{document}
\title{Interactive-Automatic Segmentation and Modelling of the Mitral Valve}

\author{Patrick Carnahan\inst{1,2}\Letter \and
Olivia Ginty\inst{2} \and
John Moore\inst{2} \and
Andras Lasso\inst{3} \and
Matthew A. Jolley\inst{4,5} \and
Christian Herz\inst{4} \and
Mehdi Eskandari\inst{6} \and
Daniel Bainbridge\inst{7} \and
Terry M. Peters\inst{1,2,8}}
\authorrunning{P. Carnahan et al.}
%
\institute{School of Biomedical Engineering, Western University, London, ON, CA\\
\email{pcarnah@uwo.ca}
\and
Imaging, Robarts Research Institute, Western University, London, ON, CA\\
\and
Laboratory for Percutaneous Surgery, Queen's University, Kingston, ON, CA\\
\and
Department of Anesthesiology and Critical Care Medicine, Children's Hospital of Philadelphia, Philadelphia, PA, USA
\and
Division of Pediatric Cardiology, Children's Hospital of Philadelphia, Philadelphia, PA, USA
\and
King's College Hospital, Denmark Hill, London, UK
\and
Department of Anesthesiology, London Health Sciences Centre, Western University, London, ON, CA
\and
Departments of Medical Biophysics, Medical Imaging, Western University, London, ON, CA\\
}

\maketitle              
\begin{abstract}
Mitral valve regurgitation is the most common valvular disease, affecting 10\% of the population over 75 years old. Left untreated, patients with mitral valve regurgitation can suffer declining cardiac health until cardiac failure and death. Mitral valve repair is generally preferred over valve replacement. However, there is a direct correlation between the volume of cases performed and surgical outcomes, therefore there is a demand for the ability of surgeons to practice repairs on patient specific models in advance of surgery. This work demonstrates a semi-automated segmentation method to enable fast and accurate modelling of the mitral valve that captures patient-specific valve geometry. This modelling approach utilizes 3D active contours in a user-in-the-loop system which segments first the atrial blood pool, then the mitral leaflets. In a group of 15 mitral valve repair patients, valve segmentation and modelling attains an overall accuracy (mean absolute surface distance) of \SI{1.40 \pm 0.26}{\milli\metre}, and an accuracy of \SI{1.01 \pm 0.13}{\milli\metre} when only comparing the extracted leaflet surface proximal to the ultrasound probe. Thus this image-based segmentation tool has the potential to improve the workflow for extracting patient-specific mitral valve geometry for 3D modelling of the valve.
\vspace{-1em}
\vspace{-1em}
\end{abstract}

\subsubsection{Acknowledgements}

The authors would like to acknowledge the following sources of funding: the Canadian Foundation for Innovation (20994), the Canadian Institutes for Health Research (FDN 201409), and the Natural Sciences and Engineering Research Council of Canada (RGPIN 2014 04504).

\vspace{-1em}

\section{Introduction}

The mitral valve is an anatomically complex, dynamic structure integral for efficient blood flow and therefore healthy cardiac output. When it becomes dysfunctional, referred to as mitral regurgitation, patients can face declining cardiovascular health until cardiac failure and death, in the absence of intervention. Furthermore, mitral regurgitation is the most common valvular disease affecting approximately 10\% of those over 75 years old \cite{Benjamin2018}. The preferred intervention for mitral regurgitation is repair, due to superior patient outcomes compared to replacement \cite{Ailawadi2008,McNeely2015}. However, the repair must be tailored to the patient-specific anatomy and pathology, which requires expert training and experience. Consequently, there is a need for patient-specific models that can permit the training and procedure-planning of patient-specific repairs to minimize its learning curve and preventable errors \cite{Eleid2016,Holzhey2013}. Previous work has demonstrated the potential for patient-specific valve modelling for both surgical training as well as preoperatively predicting surgical outcomes \cite{Ginty2018}. The cause of mitral regurgitation varies across patients, as any failure of these structures can render the valve inefficient. Therefore, segmenting the mitral valve faces both the challenge of capturing anatomy that is complex, and in motion.

In order to prepare patient-specific models, the mitral valve must first be extracted from patient image data. There are several challenges specific to mitral leaflet segmentation in 3D TEE images. There is no intensity-based boundary between leaflets and adjacent heart tissue, and distinguishing between the anterior and posterior leaflets in the coaptation zone during systole is difficult due to the lack of an intensity-based boundary. Additionally, in diastole there can be signal dropout which appears as gaps in the leaflets. To facilitate clinical use and repeat-ability, several mitral leaflet segmentation methods have been proposed. These methods focus on varying goals between deriving quantitative valve measurements and extracting annular and leaflet geometry from 3D transesophageal echocardiograph (TEE) images. Burlina et al. proposed a semi-automatic segmentation method based on active contours and thin tissue detection for the purpose of computational modelling \cite{Burlina2010}. Scheinder et al. proposed a semi-automatic method for segmenting the mitral leaflets in 3D TEE over all phases of the cardiac cycle \cite{Schneider2011}. This method utilizes geometric priors and assumptions about the mechanical properties of the valve to model the leaflets through coaptation with a reported surface error of \SI{0.84}{\milli\metre}. However, this method only represents the mitral leaflets as a single medial surface, rather than structures with thickness. Additionally, several fully automatic methods have been proposed that are based on population average atlases. Ionasec et al. describe a technique which uses a large database of manually labelled images and machine learning algorithms to locate and track valve landmarks \cite{Ionasec2010}. While this method is fully automatic, the use of sparse landmarks potentially limits the patient-specific detail that can be extracted. Pouch et al. also describe a fully automatic method which utilizes a set of atlases to generate a deformable template which is then guided to the leaflet geometry using joint label fusion \cite{Pouch2014}. The surface error of this method is reported at 0.7mm, however this is only achieved on healthy valves and performance is reduced when segmenting diseased valves. This method also may be limited by the quality of the input atlases, and could be biased towards the atlas geometry, limiting its potential for patient specific modelling. This is especially true considering the wide variety of pathologies and corresponding variety of morphologies possible with the MV.

Automatic 3D segmentation methods offer significant implications for the feasibility of patient-specific modelling in clinical use. Existing methods have demonstrated the ability to accurately segment the mitral valve structure, however remain highly time-intensive. Furthermore, some of these published methods show decreased performance when applied to highly diseased valves, demonstrating limitations in patient-specificity.

We aim to develop a segmentation method that can be applied to both normal and highly diseased valves, to extract patient-specific leaflet geometry. Our focus is on delineating the leaflet surfaces for the purpose of creating molds for our related patient-specific MV modelling project, where silicone is applied to the molds to create valves for use in surgical training and planning \cite{Ginty2018}. To accomplish this, we propose a semi-automatic segmentation method based on active contours that iterates using a user-in-the-loop strategy.

\section{Methods}

\subsection{Image Acquisition and Data Sets}

Fifteen patients with mitral valve regurgitation undergoing cardiac surgery were imaged preoperatively using Philips Epiq and iE33 systems as per clinical protocol. Of the fifteen patient datasets, six were acquired at King's College Hospital, London, UK, from patients with severe mitral regurgitation and nine were acquired at University Hospital, London, Canada. The 3D TEE images were exported into Cartesian format. The SlicerHeart module was used to import the Cartesian DICOM files into 3D Slicer \cite{Scanlan_2017}. Images at end-diastole were selected for image analysis. The exported Cartesian format images have an axial resolution of approximately \SI{0.6}{\milli\metre}.

\subsection{Semi-Automated Image Analysis}

Our software has been developed in the 3D Slicer\footnote{\url{www.slicer.org}} platform and utilizes the The Insight Segmentation and Registration Toolkit (ITK)\footnote{\url{www.itk.org}} software package. The iterative steps in our method do not use a fixed stopping point to account for the high variability in TEE data. Instead, the user runs the segmentation steps in increments until they are satisfied with the results. This leads to an ideal compromise between human judgement in ambiguous cases and guided automatic segmentation for ease of use and time efficiency. In addition, a user can view the result of the next step of the segmentation, compare it to the previous step, and make manual adjustments between active contour steps. We base our segmentation on the end diastole image where the leaflets are least likely to experience signal dropout, but where the anterior and posterior boundary is still clearly identifiable.

Before beginning the segmentation process, the user must define the valve annulus by placing a series of points. This is accomplished through the SlicerHeart software, which facilitates the placing of the points and fits a smooth annulus curve \cite{Nguyen2019}. This annulus definition is used throughout the automated process to provide context for the valve center, orientation and boundaries.

\begin{figure}[b]
    \centering
    \begin{tabular}{c c c}
        \includegraphics[width=.27\textwidth]{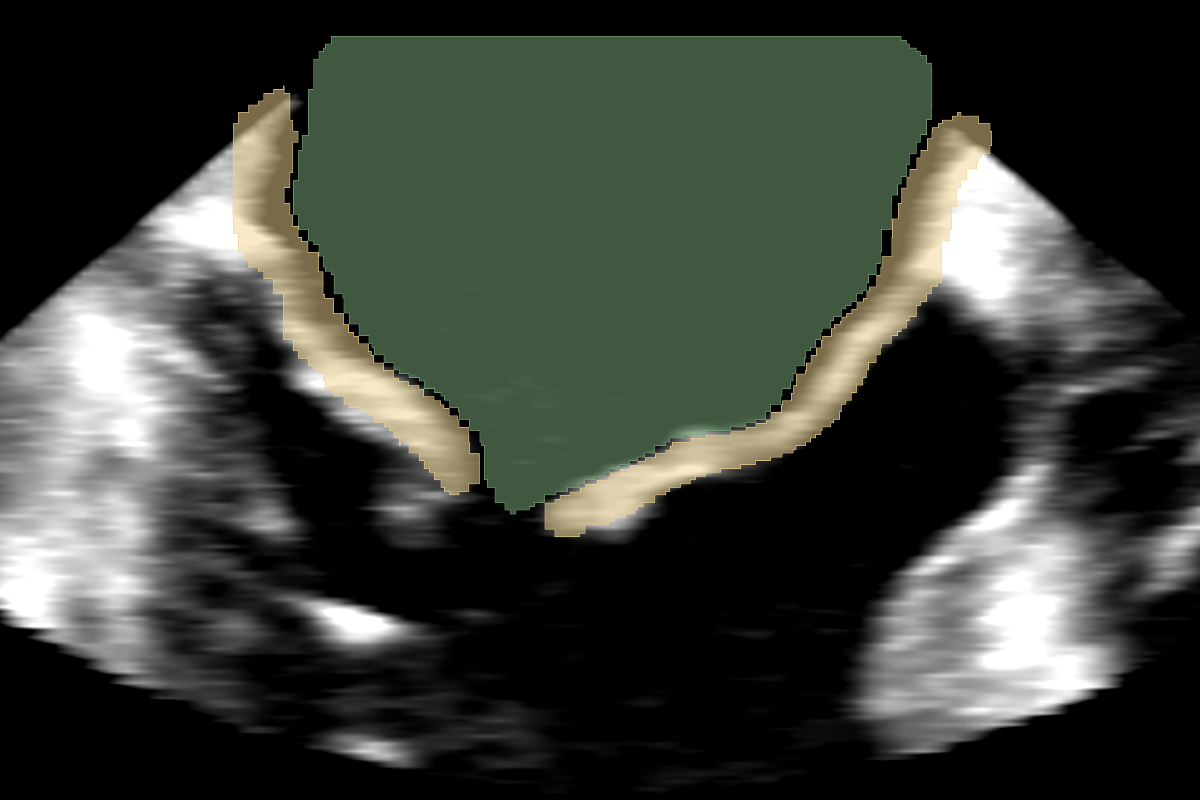} & \includegraphics[width=.27\textwidth]{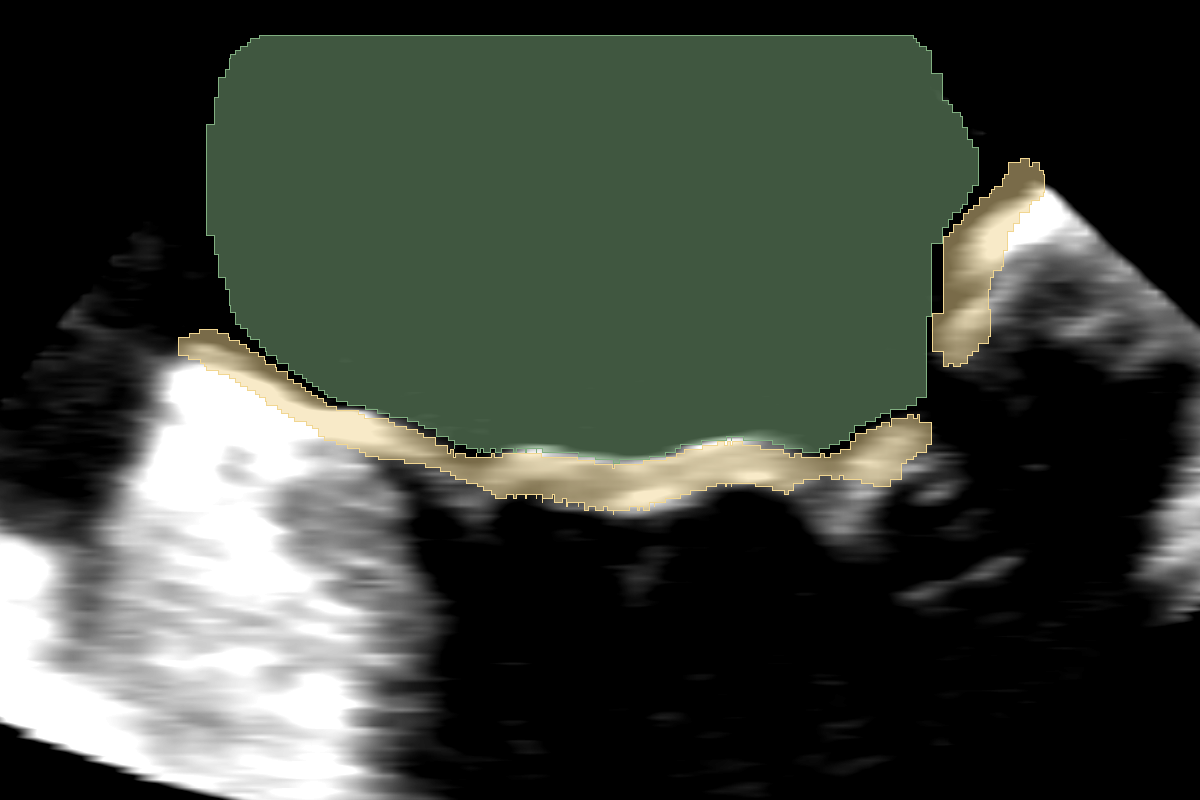} &
        \includegraphics[width=.27\textwidth]{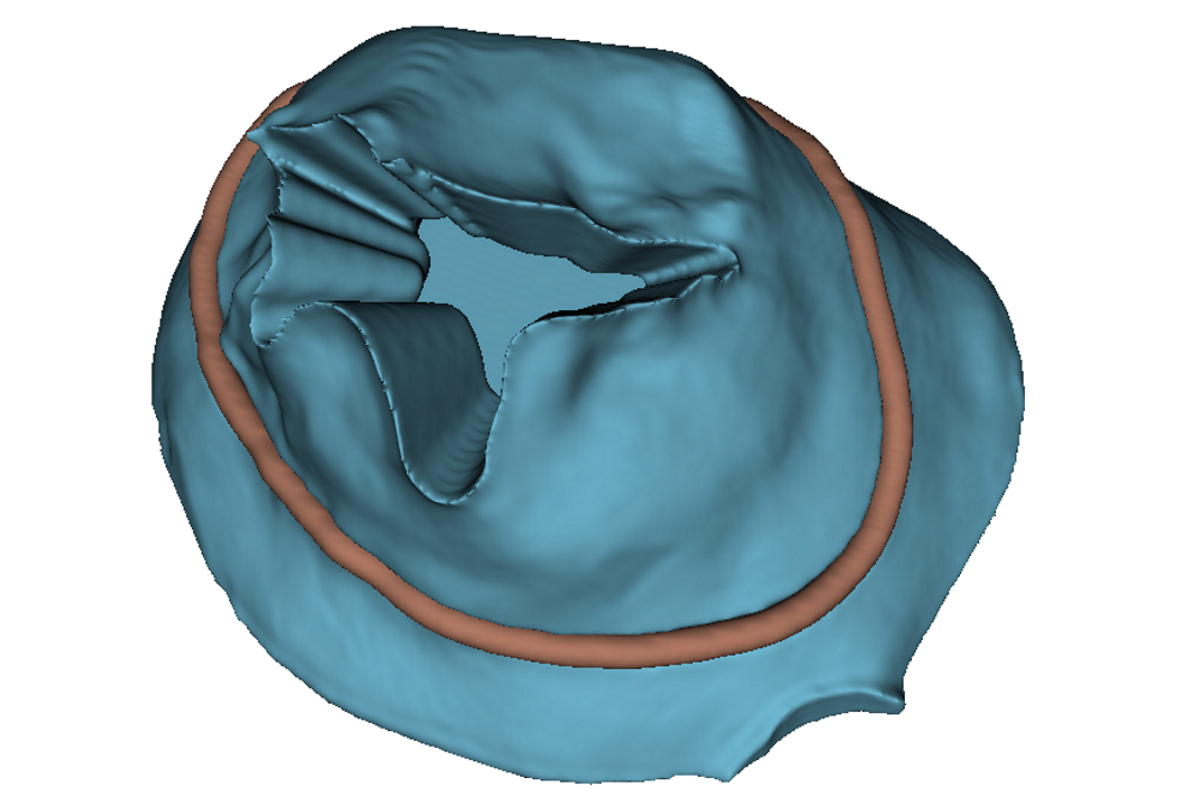} \cr
        (a) & (b) & (c)
    \end{tabular}
    \caption{Cross sectional views of a 3D TEE image and segmentation (a,b), with the blood pool segmentation shown in green and the leaflet segmentation shown in yellow. Rendering of extracted proximal surface mold (c).}
    \label{fig:segExFig}
\end{figure}

\vspace{-1em}

\subsubsection{Blood Pool Segmentation}

We first segment the atrial blood pool (BP). This blood pool segmentation provides context for the leaflet segmentation as well as the leaflet surface extraction. The image is first processed using a Gaussian filter with a variance of \SI{1}{\milli\metre}, followed by a gradient magnitude filter. This creates a feature image highlighting the contrast edges, which determines the speed of the active contour growth. The center of the defined annulus is used to initialize a geodesic active-contour filter from ITK that grows to complete the blood pool segmentation, as pictured in \autoref{fig:segExFig}. The active contour process is run with curvature, advection, and propagation scaling parameters of 1.2, 1.0, and 0.9 respectively. The curvature parameter controls boundary smoothing, while the advection parameter influences attraction to edges. The propagation scaling parameter applies an inwards or outwards force on the contour boundary creating a bias to either grow or shrink.

\vspace{-1em}

\subsubsection{Shrinking Leaflet Segmentation}

The boundary region of the blood pool segmentation within a distance of 11 voxels, or \SI{\sim 5}{\milli\metre} is taken as the initial estimate for the leaflet segmentation. This estimate is refined using another active contour approach which shrinks the segmentation down to the desired result. The active contour approach used here differs from the one used in the blood pool segmentation mainly in that it is biased to shrink. The parameters used for this phase are 0.9, 0.1, and -0.4 for curvature, advection and propagation scaling respectively. As the active contour process iterates, the segmentation pulls back to the leaflet boundaries as pictured in \autoref{fig:segExFig}. Since our approach is interactive, the user is able to view adjacent image frames during the segmentation process to better inform their decision on the ideal stopping point. In addition, a volume rendered view can also be displayed alongside a 3D mesh of the segmentation, again providing more information to the user for completing the guided segmentation.

\subsubsection{Proximal Surface Extraction}

For manufacturing our physical MV models as described in previous work \cite{Ginty2018}, we require a geometric model of the valve surface proximal to the TEE transducer. The proximal surface is extracted from the leaflet segmentation using the defined annulus for context. For points above the annulus plane, the surface is extracted by checking for self-intersection of the line between the annulus center and the point of the surface of the segmentation. This process finds all points on the inner surface, while excluding those on the outer surface. For points below the annulus, which includes the atrial wall, we compare the normal vector of the points on the leaflet segmentation surface and the corresponding closest point on the BP segmentation. Only points where the angle between the normal exceeds \SI{100}{\degree} are kept. When combined, we are left with the inner surface of the valve segmentation, which can then be used to create a mold for 3D printing pictured in \autoref{fig:segExFig}.

\section{Evaluation and Results}

In order to evaluate the accuracy of the proposed segmentation method, we compared automated segmentations to expert manual segmentations for images from 15 subjects. The ground truth expert segmentations were created using manual segmentations performed by two clinical users. Our semi-automatic system was then used on the same images, and same reference frames. Our system was used with no manual user intervention between iterations in order to perform a baseline assessment of the algorithm independent of manual influence. Comparisons were made using the mean absolute surface distance (MASD) between the boundaries of the complete segmentations, as well as the MASD between the extracted proximal surfaces.

The results indicate a MASD for the proximal surface of \SI{1.01 \pm 0.13}{\milli\metre}, which is on the order of one to two image voxels along the depth of the image. The MASD for the complete valve boundary is higher at \SI{1.40 \pm 0.26}{\milli\metre}, or two to three image voxels. The maximum local error observed was \SI{9.4}{\milli\metre} occurring at the boundary between the leaflet and the surrounding atrial tissue.  The average completion time using the proposed semi-automated segmentation method was \SI{8.93 \pm 2.31}{\minute}, compared to the manual segmentation times which averaged \SI{55.84 \pm 12.87}{\minute}. There was an overall average speedup of \SI{46.47 \pm 10.64}{\minute} using the semi-automated method over performing manual tracings.

\begin{figure}[htb]
    \centering
    \includegraphics[width=0.85\textwidth]{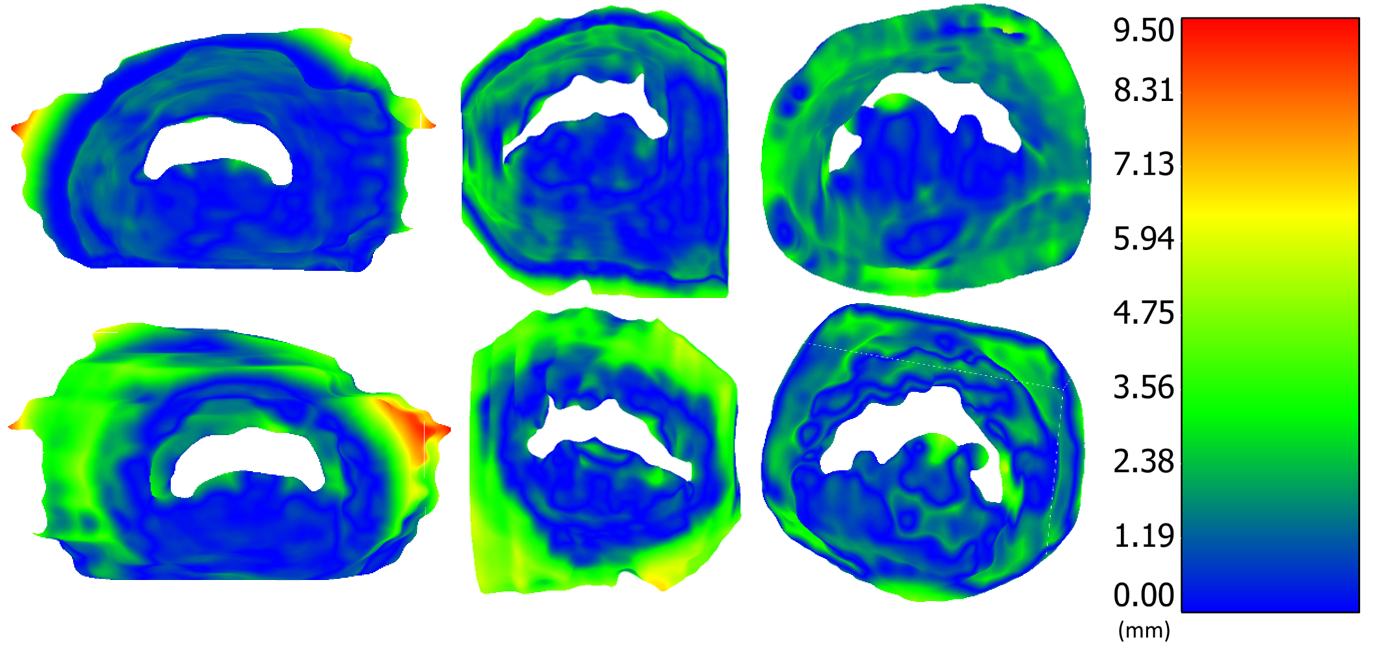}
    \caption{Distance maps of the proximal segmentation surface (top) and the distal segmentation surface (bottom) for 3 cases. The largest regions of error are located where the leaflets meet the atrial wall, as well as near the valve commisures.}
    \label{fig:proxDistFig}
\end{figure}

\section{Discussion}

The proposed semi-automatic guided segmentation method enables the extraction of mitral leaflet geometry from 3D TEE in a 3D printable form. This method allows a user to rapidly segment the mitral valve at end-diastole to extract its atrial surface for generating a patient-specific mold with minimal effort. The interactive, iterative nature of this segmentation system allows it to be used on a wide variety of pathological valves, as well as consistently work with a large range of image qualities from different systems.

Our results indicate similar overall performance to other semi-automatic methods and is on the same magnitude as previously reported inter observer variability of \SI{0.6\pm0.17}{\milli\metre} \cite{Jassar2011}. Our results are also consistent with previous studies which have observed that the greatest variability in manual and automatic segmentations occur at the boundaries of the model, rather than at the leaflet surfaces. As \autoref{fig:proxDistFig} shows, we see a boundary displacement of 1mm or less for most of the leaflet surface, and see larger displacement only at the boundary of the leaflet as well as the chordae attachment points. This is a result of the somewhat arbitrary, non-contrast boundary between the leaflets, the atrial wall, and the surrounding tissue. In addition, our method shows improved results when only considering the atrial valve surface, the surface of the valve more proximal to the TEE probe. This region shows the best image contrast and is consequently the most consistently identified between manual and automatic segmentations. Furthermore, since we do not rely on any prior data or image atlases, our system does not demonstrate any biases or drop in performance for previously unseen valve geometries. This is a critical consideration when modelling highly diseased valves for preoperative planning, as a wide variety of valve geometries can be observed. For the purposes of patient-specific valve modelling, prior work has demonstrated the creation of physical valve models using silicone and 3D printed molds based on the proximal valve surface \cite{Ginty2018}. Our method is well suited to the task of extracting mitral valve geometry for the creation of patient-specific models such as these.

\subsection{Limitations and Future Work}
 In cases of very poor image quality, the automated segmentation may miss regions of dropout, or fail to capture the exact valve geometry. Expert users performing these same segmentations often use other points in the cardiac cycle to inform their segmentation. Adapting this software to utilize additional time points may help to improve the robustness of the method in very difficult cases. In addition, color Doppler images are often captured in order to diagnose mitral valve regurgitation. This additional information could be used to increase the accuracy of segmentations as it contains information about the flow of blood between the leaflets. Further work validating the effectiveness of this system may be necessary to ensure it accurately captures patient-specific detail across healthy, mildly diseased and severely diseased valves. Our semi-automatic segmentation technique is currently being used as part of an ongoing prospective study which aims to validate the effectiveness of our dynamic silicone based valves for predicting surgical outcomes.

\section{Conclusion}

We present a technique that enables the extraction of mitral leaflet geometry from 3D TEE for  creating 3D printed models used in creating accurate patient specific physical models. Segmentations from our software successfully replicate gold-standard MV segmentations within reasonable tolerance with respect to image resolution. The overall mean surface distance analysis demonstrates that our software can extract the proximal surface of the MV to within approximately one millimeter. This level of accuracy of the is suitable for patient-specific mitral valve modelling applications. This segmentation software reduces the time required for completing an accurate mitral valve segmentation and improves the workflow of the mitral valve modelling process.

%
%

%
\end{document}